\documentclass[%
 aip,
 amsmath,amssymb,
 reprint,%
]{revtex4-1}
\usepackage{graphicx}
\usepackage{dcolumn}
\usepackage{bm}

\usepackage[utf8]{inputenc}
\usepackage[T1]{fontenc}
\usepackage{etoolbox}
\usepackage[dvipsnames]{xcolor}
\usepackage{amsmath}
\usepackage{txfonts} 

\makeatletter
\def\@email#1#2{%
 \endgroup
 \patchcmd{\titleblock@produce}
  {\frontmatter@RRAPformat}
  {\frontmatter@RRAPformat{\produce@RRAP{*#1\href{mailto:#2}{#2}}}\frontmatter@RRAPformat}
  {}{}
}%
\makeatother
\begin{document}

\preprint{AIP/123-QED}

\title{Frustrated hops in Ring Polymer Surface Hopping: Real-time dynamics and detailed balance}

\author{Dil K. Limbu}
\author{Farnaz A. Shakib}
 \email{shakib@njit.edu.edu.}
\affiliation{ 
Department of Chemistry and Environmental Science, New Jersey Institute of Technology, Newark 07102, NJ United States
}

\date{\today}

\begin{abstract}
Ring Polymer Surface-Hopping (RPSH) has been recently introduced as a well-tailored method for incorporating nuclear quantum effects (NQEs), such as zero-point energy and tunneling, into non-adiabatic molecular dynamics simulations. The practical widespread usage of RPSH demands a comprehensive benchmarking of different reaction regimes and conditions with equal emphasis on demonstrating both the cons and pros of the method. Here, we investigate the fundamental questions related to the conservation of energy and detailed balance in the context of RPSH. Using Tully's avoided crossing model as well as a 2-level system coupled to a classical bath undergoing Langevin dynamics, we probe the critical problem of the proper treatment of the classically forbidden transitions stemming from the surface hopping algorithm. We show that proper treatment of these frustrated hops is key to the accurate description of real-time dynamics as well as reproducing the exact quantum Boltzmann population. 
\end{abstract}

\maketitle

\section{\label{sec:level1}Introduction}
Trajectory-based mixed quantum-classical dynamics methods have been a mainstream approach in describing electronic and/or nuclear quantum effects in condensed-phase dynamics.\cite{Brown:2021,Crespo-Otero:2018,Wang:2016,Tully:2012,Barbatti:2011} For more than three decades, Tully's Fewest Switches Surface Hopping (FSSH)\cite{Tully:1990} has served as ground zero for developing a myriad of such methods. In the original FSSH, the classical degrees of freedom (DOFs) evolve on the adiabatic surfaces interrupted by instantaneous electronic transitions. For each nuclear trajectory, the associated electronic density matrix is determined by coherently integrating the electronic time-dependent Schr\"{o}dinger equation along the classical trajectory. The transition probabilities between adiabatic states are then determined from the density matrix elements and their time derivatives. Hammes-Schiffer and Tully extended this methodology to non-adiabatic transitions between vibrational states as well, which allowed simulation of proton-transfer (PT)\cite{Hammes-Schiffer:1994} and proton-coupled electron transfer (PCET)\cite{Fang:1997a,Fang:1997b,Fang:1999} reactions. Efforts to remedy the primary limitation of the FSSH ansatz, commonly known as overcoherence problem or lack of decoherence, gave birth to more sophisticated trajectory-based methods such as augmented surface hopping,\cite{Jain:2016} decoherence-induced surface hopping,\cite{Jaeger:2012} simultaneous-trajectory surface hopping,\cite{Shenvi:2011} and coherent fewest-switches surface-hopping,\cite{Tempelaar:2018} to name a few. The lack of a formal derivation of surface hopping was met by the advent of a first-principle trajectory-based methodology utilizing quantum-classical Liouville equation (QCLE)\cite{Kapral:1999,Nielsen:2000} which was proved to be very accurate for PT\cite{Hanna:2005} reactions. Shakib and Hanna extended QCLE to the study of the rate and mechanism of PCET\cite{Shakib:2014,Shakib:2016a,Shakib:2016b}  reactions yielding exceptionally accurate results, albeit at the expense of a huge computational cost. The number of classical trajectories to reach convergence in the QCLE approach for PCET reactions turned out to be at the order of millions,\cite{Shakib:2016b} which is 1000 times more than the FSSH requirement. Nevertheless, later, Kapral showed that by applying two approximations, surface-hopping could be derived from QCLE\cite{Kapral:2016} hence putting the former on much firmer ground than the previously phenomenologically perceived one. Very recently, developments aimed at dealing with the limitations of FSSH \textit{via} a mapping approach to surface hopping (MASH)\cite{Mannouch:2023}, which imposes internal consistency between nuclear and electronic DOFs. Still, MASH performs its best with the addition of decoherence; however, this is not an \textit{ad hoc} correction scheme but the result of the rigorous derivation of the method at the QCLE limit.\cite{Mannouch:2023} In 2012, Shushkov \textit{et al.} tried to remedy another inherent limitation of surface-hopping,\cite{Shushkov:2012} i.e., lack of nuclear quantum effects (NQEs), by marriage between surface hopping and ring polymer molecular dynamics (RPMD)\cite{Craig:2004}. In this method, the non-adiabatic electronic transitions are described by the FSSH algorithm. At the same time, NQEs are incorporated with Feynman's imaginary-time path-integral formalism,\cite{Feynman:1965} giving birth to ring polymer surface hopping (RPSH). Thus, RPSH is deemed a well-tailored method for investigating multi-electron/multi-proton transfer dynamics in condensed phases. Shakib and Huo applied RPSH with centroid approximation to investigate electronic non-adiabatic dynamics in three infamous Tully models with explicit nuclear quantization.\cite{Shakib:2017} They showed that RPSH can qualitatively capture the correct branching probabilities, especially at low-temperature limits, due to the inclusion of nuclear tunneling and zero-point energy via the extended phase-space of the classical ring polymer. Interestingly, RPSH was also capable of quantitatively reproducing the branching probabilities in the model of "extended coupling with reflections", which is specifically designed for the study of coherence/decoherence. 

Despite all the promises that RPSH offers, it is still an approximate method whose accuracy and validity limits should be carefully examined. Specifically, while we know that it combines the powerful capabilities of FSSH and RPMD, we should also ask about the shortcomings of its constituent methodologies. Are they magnified or quenched within RPSH? And to what degree? In the current work, we address the critical issues related to surface hopping phenomenon. We will focus on the conservation of energy, which leads to frustrated hops, and their effect on preserving detailed balance. We show how these frustrated hops are manifested in RPSH and offer different remedies for dealing with them. The effect of such remedies is discussed in terms of different model systems to provide valuable insights into the choice of proper recipe for treating frustrated hops with respect to the objectives of a study. While not a single treatment of frustrated hops seems applicable in all models, RPSH systematically operates better than FSSH. This is on top of the inclusion of NQEs into the dynamics that separates RPSH from the ever-expanding pool of surface hopping methods. Further, this paper portrays a clear picture of the functionality of RPSH, which is crucial for future developments \textit{via} a unified community effort. In the remainder of this paper, we will first give a summary of the RPSH algorithm at the centroid level as well as methodological considerations regarding frustrated hops. The studied models will be introduced at the beginning of each subsection in the Results and Discussions. Concluding Remarks and our Outlook toward future research directions in this field will follow.

\section{Methodological Considerations} 
\subsection{\label{sec:RPSH}Ring Polymer Surface Hopping}
We present a brief introduction to the RPSH ansatz with centroid approximation, interested reader is referred to Refs.~\citenum{Shushkov:2012} and~\citenum{Shakib:2017} for further details. In RPSH, every nuclear DOF is represented by a ring polymer which is comprised of \textit{n} copies of the nuclear DOF, known as beads, connected by harmonic forces. The corresponding extended Hamiltonian is described with:
\begin{equation}
 H_n=\sum^n_{i=1}\left[\frac{\textbf{P}_i^2}{2M}+\frac{M\omega_n^2}{2}(\textbf{R}_i-\textbf{R}_{i-1})^2+V_{\alpha}(\textbf{R}_i)\right]
\label{Hamiltonian}.   
\end{equation}
Here, $n$ is the total number of beads and $\omega_n=n/\beta\hbar$ where $\beta=1/k_{\rm{B}}T$ is the reciprocal temperature. $\textbf{P}_i$ and $\textbf{R}_i$ represent the momentum and position of each bead of the ring polymer which moves on a single adiabatic surface $|\alpha;\textbf{R}_i\rangle$ corresponding to the potential energy $V_{\alpha}(\textbf{R}_i)=\langle\alpha;\textbf{R}_i|\hat{V}|\alpha;\textbf{R}_i\rangle$. At every time step of the nuclear dynamics, the position and momentum of the centroid of the ring polymer is updated as:
\begin{equation}
\bar{\textbf{R}}=\frac{1}{n}\sum_{i=1}^n\textbf{R}_i\;\;\;\;\;\;\;\;\;\;\;\;\;    \bar{\textbf{P}}=\frac{1}{n}\sum_{i=1}^n\textbf{P}_i   
\end{equation}
At this point, according to the surface hopping algorithm, the time-dependent Shr\"odinger equation is numerically integrated as
\begin{equation}
i \hbar \dot{c}_{\alpha}(t)=V_{\alpha}(\bar{\textbf{R}})\,c_{\alpha}(t)-i\hbar\sum_{\beta}\dot{\bar{\textbf{R}}}\cdot \textbf{d}_{\alpha\beta}(\bar{\textbf{R}})\,c_{\beta}(t)   \label{TDSE} 
\end{equation}
to confer the electronic coefficients $c_{\alpha}$ associated with each adiabatic surface. The important difference with the original FSSH is that this integration is carried out along the motion of the centroid. Both the energy of the adiabatic surfaces, $V_{\alpha}(\bar{\textbf{R}})$ and the non-adiabatic coupling vector between surfaces, $\textbf{d}_{\alpha\beta}(\bar{\textbf{R}})=\langle\alpha;\bar{\textbf{R}}|\nabla_{\bar{\textbf{R}}}|\beta;\bar{\textbf{R}}\rangle$, are evaluated at the centroid level. The same goes for the probability of transition, i.e., switching between surfaces at each time step $\Delta t$, which is defined based on density matrix elements, $\rho_{\alpha\beta}=c_{\alpha}c_{\beta}^{*}$, as
\begin{equation}
 g_{\alpha\beta}=\frac{-2\mathrm{Re}(\rho_{\beta\alpha}^{*}\dot{\bar{\textbf{R}}}\cdot \textbf{d}_{\beta\alpha}(\bar{\textbf{R}}))\,\Delta t}{\rho_{\alpha\alpha}}.   
\end{equation}
The nonadiabatic transition from the current surface $\alpha$ to the next surface $\beta$ occurs if $g_{\alpha\beta}$ is greater than a randomly-generated number between 0 and 1. If a transition occurs, the entire ring polymer hops to the next adiabatic surface while the velocity of each bead in the ring
polymer is re-scaled in order to conserve the total energy of the quantum plus classical subsystems. 

\subsection{\label{sec:FHs}Conservation of energy and frustrated hops}
By construction, surface hopping methodologies conserve total quantum plus classical energy by enforcing the hopping trajectories to have enough kinetic energy to compensate for the potential energy difference between surfaces, or states. Transition attempts that do not fulfill this requirement are deemed as \textit{frustrated hops} and return to the original states.\cite{Tully:1990}  
Proper treatment of these frustrated hops is an essential step in carrying out surface hopping simulations. Tully demonstrated frustrated hops as trajectories hitting a wall and coming back, hence the component of their velocity in the direction of nonadiabatic coupling vector needed to be reversed.\cite{Tully:1990,Hammes-Schiffer:1994} Later, M\"uller and Stock showed that \textit{not} reversing the velocity of frustrated hops leads to significantly improved results in photoinduced relaxation dynamics in comparison to reversing the velocity\cite{Muller:1997} hinting that the choice of treatment can be system-dependent. In our first implementation of RPSH,\cite{Shakib:2017} we opted to not reverse the velocity of the beads of the ring polymer in the case of frustrated hops. To assess the accuracy of the method and shed light on manifestation of frustrated hops in RPSH, here, we revisit Tully's single avoided crossing model and establish the effect of reversing or not reversing the velocity of frustrated hops on the adiabatic population transfer profile in the low to medium momentum regions where the nuclear quantum effects are paramount. Furthermore, we employ two different remedies suggested by Truhlar\cite{Jasper:2003} and Subotnik\cite{Jain:2015} for frustrated hop treatment. Jasper and Truhlar combined the features of velocity "reversing" (VR) and "not reversing" (NR) in a prescription called $\Delta V$ which, whenever a frustrated hop occurs, allows the trajectory to feel the target adiabatic state $\beta$. Accordingly, in the case of a frustrated hop in RPSH, we compare $\dot{\bar{\textbf{R}}}\cdot \textbf{d}_{\alpha\beta}$ to the component of the force in the direction of the nonadiabatic coupling vector, defined as
\begin{equation}
F_{\beta}=-\nabla V_{\beta}({\bar{\textbf{R}}})\cdot \textbf{d}_{\alpha\beta}(\bar{\textbf{R}}).    
\end{equation}
If these two quantities have the same sign, we do not reverse the velocity but will do so if they have opposite signs. On the other hand, Jain and Subotnik suggested a comparison between the components of forces from both the current and the target adiabatic states. Accordingly, in RPSH, we calculate:
\begin{equation}
\bigg(\textbf{d}_{\alpha\beta}(\bar{\textbf{R}})\cdot \big(-\nabla V_{\alpha}({\bar{\textbf{R}}})\big)\bigg) \bigg(\textbf{d}_{\alpha\beta}(\bar{\textbf{R}})\cdot \big(-\nabla V_{\beta}({\bar{\textbf{R}}})\big)\bigg)  
\end{equation}
and reverse the velocity if it is smaller than zero. Henceforth, we shall call this approach $\Delta V^2$ as it bears resemblance to the $\Delta V$ approach.

\subsection{\label{sec:detailed_balance}Frustrated hops and detailed balance}
The requirement to conserve energy leads to disruption of internal consistency despite the original promise of surface hopping algorithm, i.e., the fraction of trajectories ended up on each state is not consistent with the average quantum probabilities anymore. Even though, frustrated hops are not the Achilles heel of the surface hopping methodologies, note that transitions from higher-energy to lower-energy states are always permitted and it is the reverse that encounters rejection. As a result, presence of frustrated hops leads to an almost correct apportionment of nuclear DOFs according to the Boltzmann distribution. It was shown early-on that in a 2-state system coupled to a chain of \textit{N} classical DOFs,  the original FSSH algorithm approximately, in some reaction regimes exactly, preserves detailed balance.\cite{Parandekar:2005,Schmidt:2008} Recently, Prezhdo and coworkers revisited this study with a focus on the relationship between detailed balance and proper treatment of frustrated hops.\cite{Sifain:2016} They showed that the VR approach slightly improves the detailed balance in this model system compared to the NR approach over a high-temperature region of 350 K-2500 K. This conclusion is, of course, opposite to the earlier findings of M\"uller and Stock\cite{Muller:1997} albeit in a different system. Here, we will employ a slightly modified version of the chain model to investigate the capability of RPSH in preserving detailed balance, with different treatments of frustrated hops, compared to the original FSSH. Furthermore, we will demonstrate our results at the significantly important region of low temperatures.

\section{Results and Discussion} 
\subsection{Single-avoided crossing model: Branching probabilities}
\begin{figure}[t]
  \includegraphics[width=.45\textwidth]{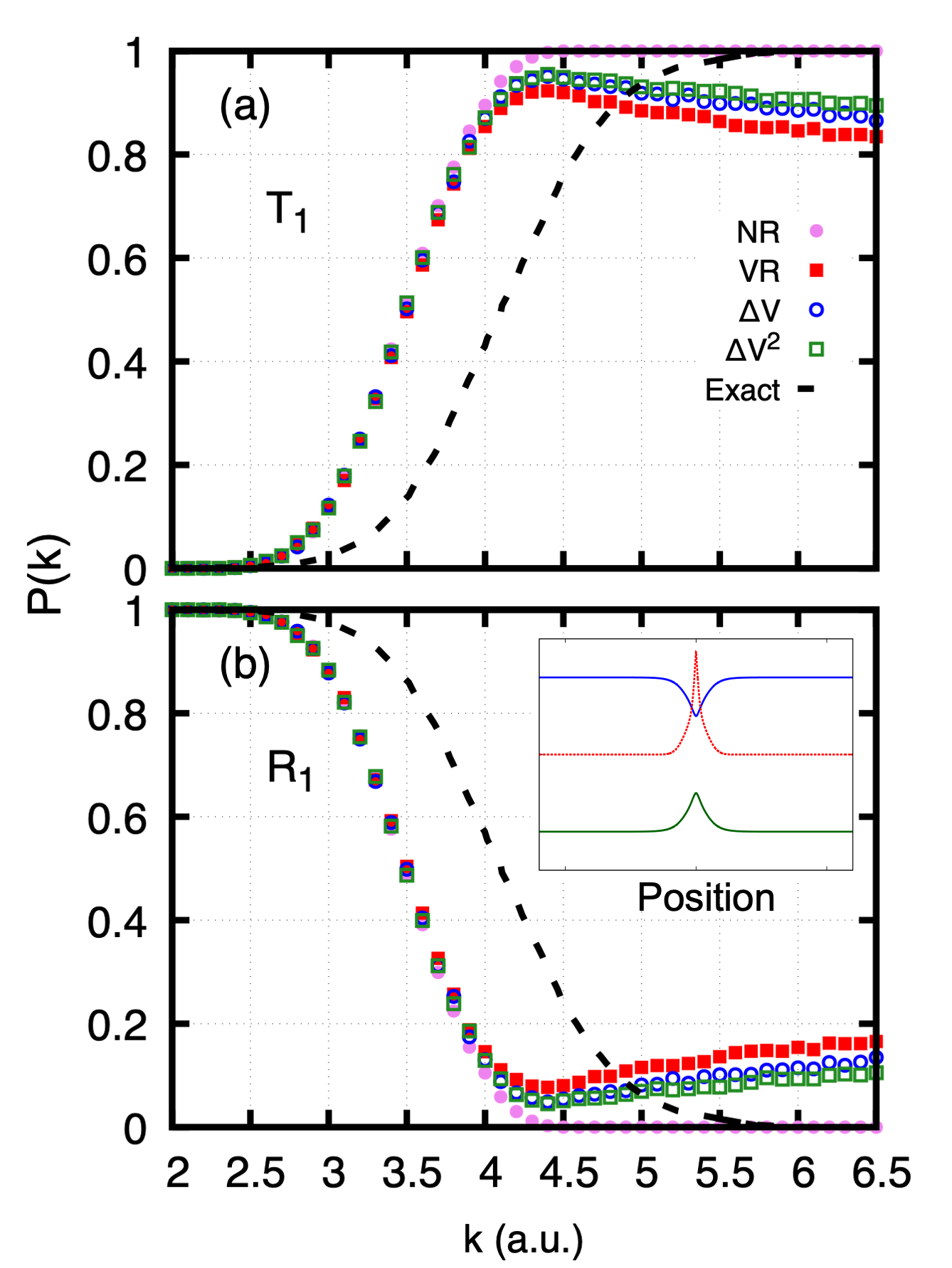}
  \caption{Branching probabilities in the Tully's single avoided crossing (SAC) model at the low momentum region; Transmission (a) and reflection (b) on state $|1\rangle$, T$_1$ and R$_1$, respectively. The model is shown in the in-set with the red line representing the non-adiabatic coupling vector.}
  \label{fig:fig2}
\end{figure}

Here, we focus on Tully's single-avoided crossing (SAC) model, shown in the inset of Fig.~\ref{fig:fig2}, which is defined by a $2\times 2$ diabatic matrix with the details being provided in the Supporting Information (SI). Diagonalizing this matrix yields two adiabatic states with a single avoided crossing and a corresponding non-adiabatic coupling vector. This 2-state system is then coupled to one nuclear DOF whose wavefunction is initialized on state $|1\rangle$ as a Gaussian wavepacket in the form of
\begin{equation}
G(R)=\bigg(\frac{2\alpha}{\pi}\bigg)^{\frac{1}{4}}\mathrm{exp}(-\alpha(R-R_0)^2+ik(R-R_0)).    
\end{equation}
This is manifested in a Gaussian distribution of nuclear position around $R_0=-15$ a.u. with the width of $\sigma_R=1/\sqrt{2\alpha}$. Here, $\alpha=0.25$ a.u. and $k$ is the deterministic incoming momentum of the nuclear DOF with the mass $M=2000$ a.u. Note that, here, the classical DOF is expanded to a ring polymer with four beads whose initial positions are randomly selected from this Gaussian distribution. The choice of four beads for the ring polymer is based on the convergence of the adiabatic populations.\cite{Shakib:2017} We first focus on the branching probabilities as a function of the incoming momentum obtained from an ensemble of 10,000 RPSH trajectories ran for 0.5 ps with a time-step of 1 a.u. ($\sim$0.024 fs). Branching probabilities show the final adiabatic populations of states $|1\rangle$ and $|2\rangle$ with differentiating between trajectories that upon reaching at avoided crossing transmit from the reactant to the product side or reflect back to the reactant side. Fig.~\ref{fig:fig2} demonstrate the branching probabilities, namely transmission and reflection on state $|1\rangle$ ($T_1$ and $R_1$, respectively), in a low-momentum region, i.e. $k=2-6.5$ a.u. 
According to the exact results shown in dashed lines, in this region, the incoming momentum is not sufficient for non-adiabatic transition and the wavepacket moves adiabatically on state $|1\rangle$.  This dynamics is governed by nuclear tunneling where exact results show a smooth increase of $T_1$ in the expense of a decrease in $R_1$.
\begin{figure}[t]
  \includegraphics[width=.45\textwidth]{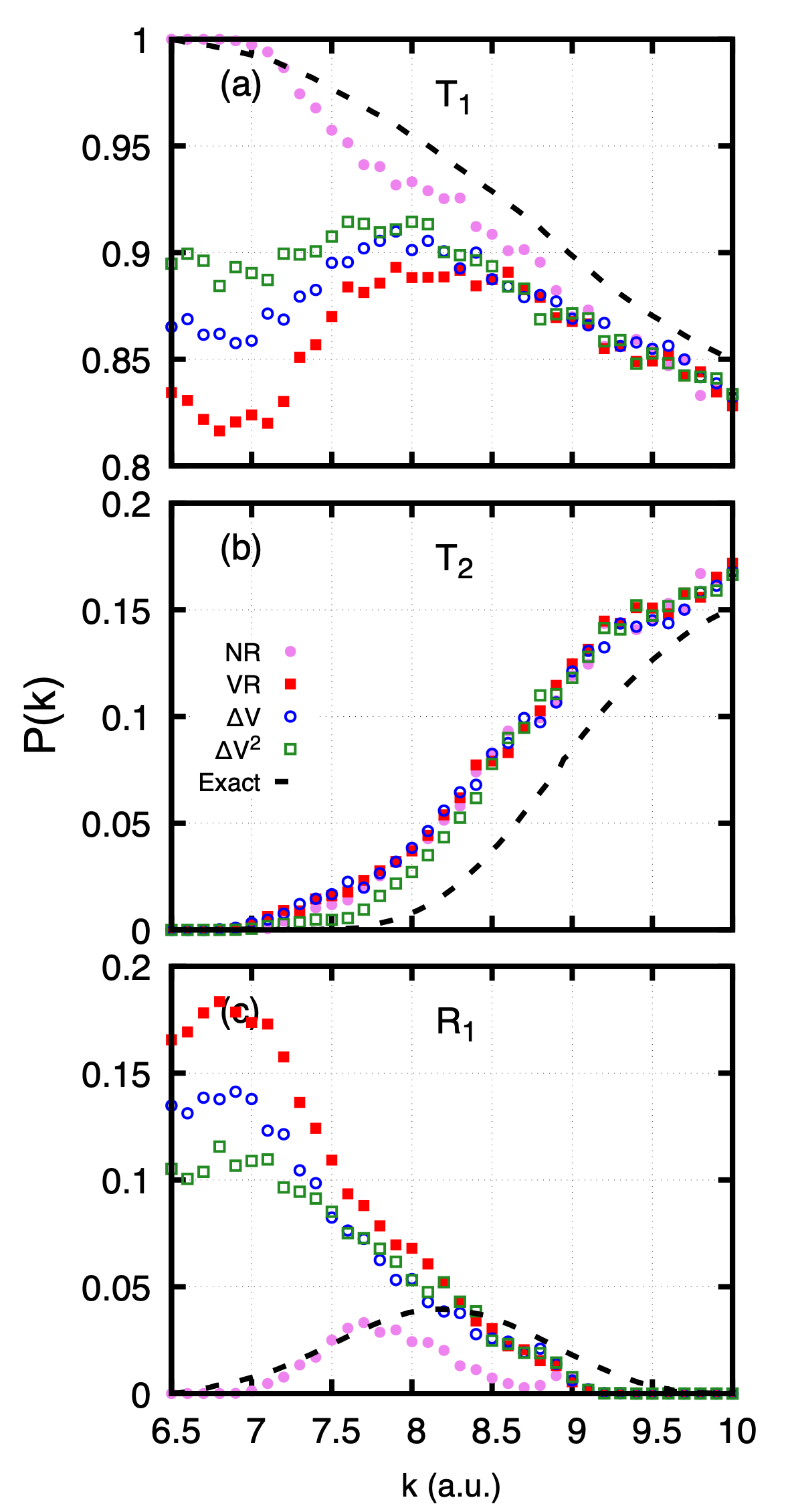}
  \caption{Branching probabilities in the SAC model at the intermediate momentum region; transmission on state $|1\rangle$, $T_1$, (a), transmission on state $|2\rangle$, $T_2$, (b), and reflection on state $|1\rangle$, $R_1$ (c).}
  \label{fig:fig4}
\end{figure}
We have previously shown\cite{Shakib:2017} how FSSH fails to capture this quantum-mechanical phenomenon but RPSH recovers the correct physical behaviour, i.e. a smooth transition between $R_1$ and $T_1$, through the extended phase-space of a classical ring polymer. Here, we evaluate the effects of the proper treatment of frustrated hops on branching probabilities applying four different schemes for adjusting their velocities, namely velocity reversing (VR) or not reversing (NR) along with $\Delta V$ and $\Delta V^{2}$ approaches\cite{Jasper:2003,Jain:2015} as explained in the previous section. As can be seen in Fig~\ref{fig:fig2}, the four schemes show similar behaviour in capturing the smooth transition in the very low momentum region but start deviating around $k=3.3$ a.u. where only the NR scheme follows the same behaviour of the exact results in 100\% reversal of $R_1$ to $T_1$. The other three schemes underestimate the magnitude of adiabatic transmission at higher values of momentum. The reason goes back to the fact that trajectories facing frustrated hops in NR scheme still continue their motion with positive momentum. They tunnel through the energy barrier on state $|1\rangle$ and end up in the product side. On the other hand, fully or partially reversing the velocity in the other three schemes forces such a trajectory to go back to the reactant side negating the possibility of tunneling through the energy barrier. This results in an artificial increase of $R_1$ in the expense of decrease in $T_1$.   
As we increase the momentum beyond $k=6.5$ a.u. we reach an intermediate reaction regime where non-adiabatic trajectories appear alongside adiabatic trajectories. This higher momentum results on successful transitions from state $|1\rangle$ to $|2\rangle$ and a smooth increase of $T_2$ as shown in Fig.~\ref{fig:fig4}b. As can be seen, all four schemes retrieve the correct physical behaviour due to the inclusion of ZPE in the dynamics by RPSH. We have already shown that FSSH fails to do so.\cite{Shakib:2017} On the other hand, there is still significant deviation between NR and the other three schemes in retrieving the correct physical behaviour in the case of $T_1$ and $R_1$ with only the former being successful.
To shed more light on this finding, we take a look at the number of frustrated hops averaged over the number of trajectories ($N_{FHs}$) as a function of the incoming momentum in the RPSH method, Fig.~\ref{fig:fig3}. A comparison between these results and Fig.~\ref{fig:fig2} and Fig.~\ref{fig:fig4} show that the deviation between the four schemes starts as soon as we have frustrated hops in our simulations, i.e. around $k=3.3$ a.u., and persists till around $k=8.9$ a.u. where $N_{FHs}$ goes to zero.
\begin{figure}[b]
  \includegraphics[width=.45\textwidth]{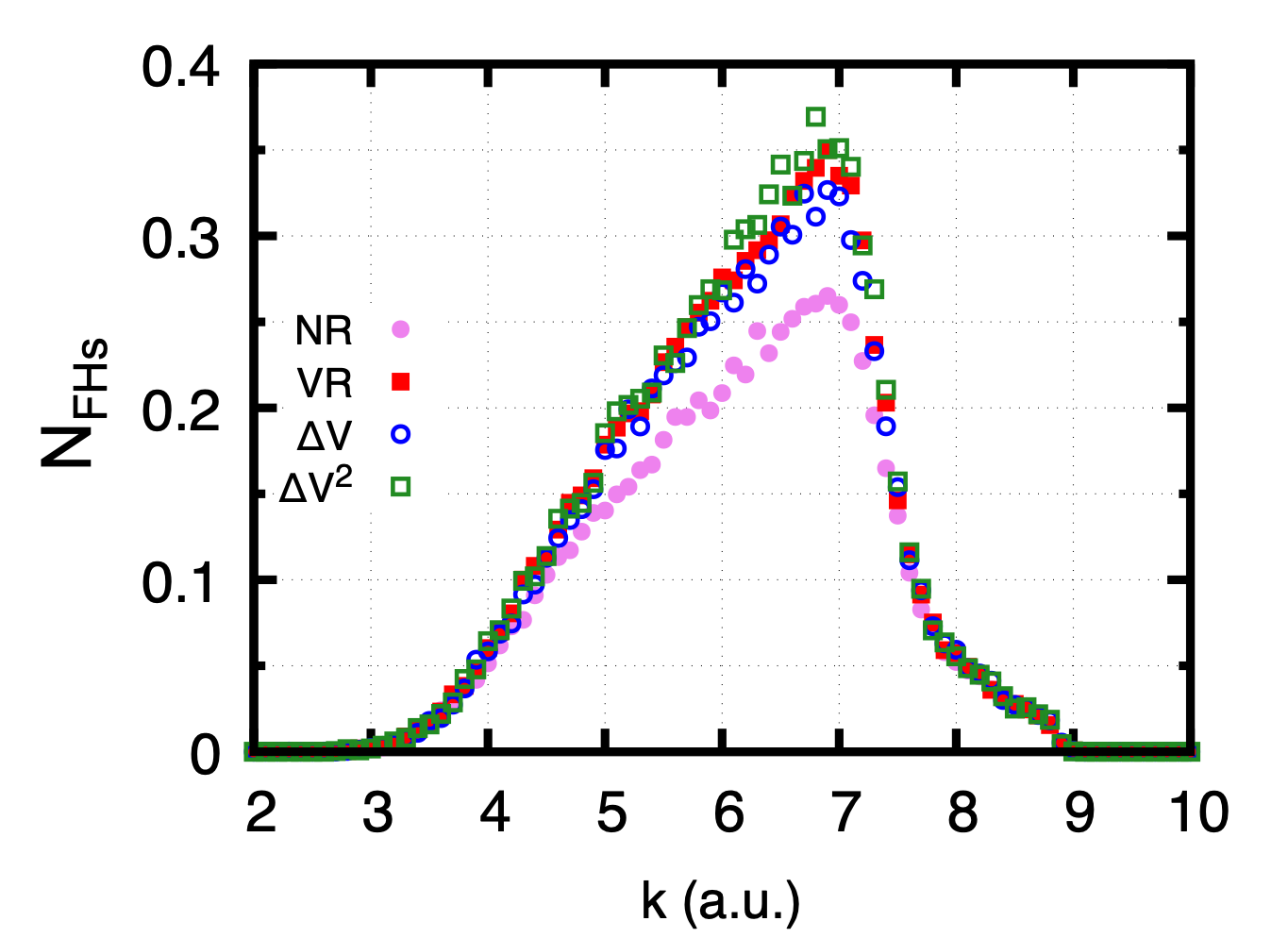}
  \caption{The number of frustrated hops at different velocity re-scaling schemes in RPSH in the low and intermediate momentum region of the SAC model.}
  \label{fig:fig3}
\end{figure}
Clearly, in the adiabatic or intermediate reaction regimes RPSH trajectories encounter frustrated hops and even partially reversing the velocity can prevent an otherwise successful trajectory dynamics from taking place. Another point seen in Fig.~\ref{fig:fig3} is the lower number of frustrated hops in the NR scheme compared to the other three schemes. The reason can be traced back to the fact that in the NR scheme, successful trajectories continue moving to the product side regardless of how many times they encounter frustrated hops or whether this encounter happens in the vicinity of an energy barrier on the reactant or the product side. For the other three schemes, on the other hand, a trajectory on the product side might encounter a frustrated hop and due to reversing the velocity tunnels back to the reactant side. Now, if that trajectory encounters another reversing of velocity it will go back to the product side. Overall, reversing the velocity allows the trajectories to spend more time on the avoided crossing region and hence $N_{FHs}$ increases. This can enhance the deviation of the results of velocity re-scaling schemes from the exact results.   

\subsection{\textit{N}-particle chain model: Detailed balance}
The original RPMD, by construction, yields real-time MD trajectories that preserve exact quantum Boltzmann distribution.\cite{Craig:2004} Its expansion to multi-state non-adiabatic dynamics, however, did not always lead to preserving the detailed balance mainly due to the zero-point energy leakage. Nevertheless, multi-variable (MV)-RPMD, where both electronic states and nuclear DOFs are represented by continuous Cartesian coordinates,\cite{Ananth:2013} demonstrated that ring polymer isomorphism can exactly preserve detailed balance in multi-state systems. On the other hand, as was explained before, FSSH  approximately preserves detailed balance due to the presence of frustrated hops. Now, the question is whether RPSH can borrow these desired characteristics and improve upon. To investigate this matter, here, we employ a 2-state system, with constant values of the energy gap ($\Delta=E_2-E_1$) and non-adiabatic coupling vector ($d_{12}$), coupled to a chain of $N$ nuclear DOFs.\cite{Parandekar:2005} 
The quantum subsystem is coupled to only the first particle in the chain and the nearest-neighbor potential energy between the particles is a quartic Morse potential as
\begin{equation}
    V(\mathbf{R}) = \sum_{i=1}^N V_M(R_i - R_{i+1})
\end{equation}
where
\begin{equation}
    V_M(R) = V_0(a^2R^2-a^3R^3+0.58a^4R^4).
\end{equation}
The atom farthest from the quantum subsystem ($R_N$) is connected to Langevin dynamics. All the particles in the chain are represented by ring polymers comprised of four beads based on our convergence tests, see the SI Figure S1. Subsequently, the EOM for the last particle would be slightly different from the conventional classical Langevin dynamics. Each mode of a free ring polymer ($k$) exhibits uncoupled harmonic oscillator dynamics when represented in the normal mode basis.\cite{Ceriotti2012} Accordingly, the corresponding EOM for the last ring polymer is written as
\begin{equation}
    \dot{P}_N^{(k)} = -\frac{\partial V}{\partial R_N^{(k)}} - \gamma^{(k)}P_N^{(k)} + \sqrt{\frac{2M\gamma^{(k)}}{\beta_n \Delta t}}.
\label{eq10}
\end{equation}
where $\beta_n=\beta/n$. 
\begin{table}[h]
\caption{\label{tab:dbparam} Simulation parameters used for the $N$-particle chain model.}
\begin{ruledtabular}
\begin{tabular}{lrl}
Parmeter & Value & Unit \\
\hline
$N$ & 20 \\ 
$m$ & 12.0 & amu \\ 
$V_0$ & 175.0 & kJ/mol  \\ 
$a$ & 4.0 & {\AA}$^{-1}$ \\ 
$\gamma$ & $10^{14}$ & $s^{-1}$ \\ 
$\Delta$ & 8.0 & kJ/mol \\
$d_{12}$ & $-$6.0 & {\AA}$^{-1}$ \\
\end{tabular}
\end{ruledtabular}
\end{table}
The second term includes the Langevin friction constant $\gamma^{(k)}$. For the excited modes of the ring polymer $(k > 0)$ $\gamma^{(k)} = 2\omega_k$ where $\omega_k = 2\sin(2\pi/n)/\beta_n\hbar$ and for the centroid mode $\gamma^{(0)} = \gamma$. The last term in Eq.~\ref{eq10} represents a random force the width of which depends both on the number of beads within $\beta_n$ and the time step of the simulation $\Delta t$. 
\begin{figure}[h!]
  \includegraphics[width=.45\textwidth]{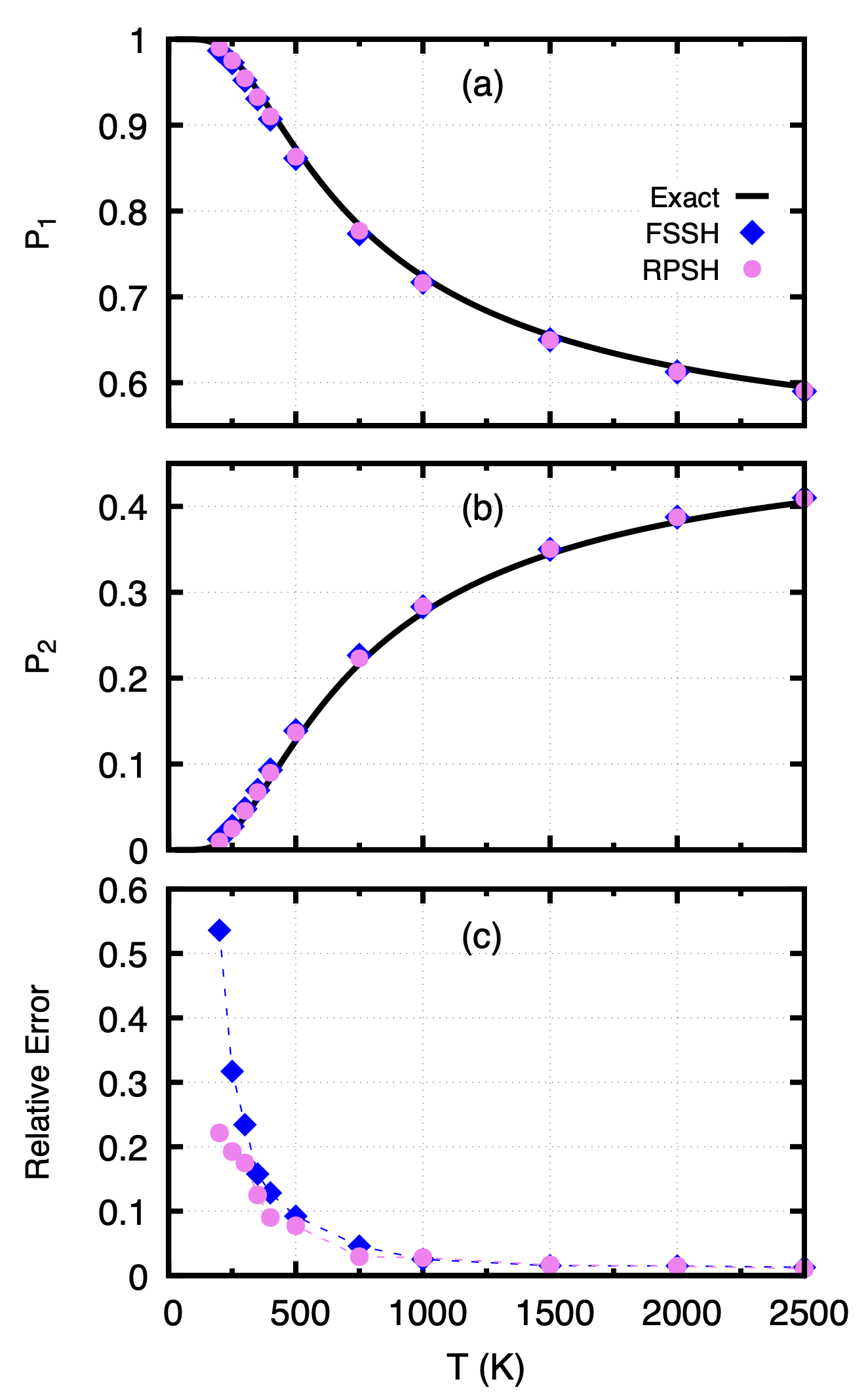}
  \caption{Equilibrium populations on the first state $P_1$ (a), and on the second state $P_2$ (b) of the 2-state quantum subsystem in the chain model obtained from RPSH ({\large\textcolor{Thistle}{$\bullet$}}) and FSSH ({\small\textcolor{blue}{$\Diamondblack$}}) methods with NR scheme. Exact Boltzmann populations are shown in solid lines. Panel (c) shows the relative errors of RPSH and FSSH in producing $P_2$ with respect to the Boltzmann population.}
  \label{fig:fig5}
\end{figure}
All parameters of our 2-state chain model are listed in Table~\ref{tab:dbparam} with some variations compared to the original model.\cite{Parandekar:2005} 
Specifically, we changed the energy gap to 8.0 kJ/mol and the nonadiabatic coupling vector to -6.0~\AA$^{-1}$ to expand our dynamics simulations to a wider range of temperatures at 200 K$-$2500 K, with a special emphasis on the low temperature region. 
Fig.~\ref{fig:fig5}a and~\ref{fig:fig5}b show the populations of the two states ($P_1$ and $P_2$) over a wide range of temperatures with the solid line representing the exact Boltzmann populations ($P_{\rm{B}}$).  The RPSH populations are obtained from 1000 trajectories for 50 ps with $\Delta t=0.01$ fs and taking the average of the last 20 ps to avoid any dependence on the initial conditions. FSSH results are obtained similarly and are shown for comparison. The RPSH results are very satisfactory but at the same time are very close to the FSSH data. For a better comparison, Fig.~\ref{fig:fig5}c demonstrates the relative error of the RPSH and FSSH results for the second state defined as $(P_{2}-P_{\rm{B}})/P_{\rm{B}}$. As can be seen, at high temperatures RPSH and FSSH show identical results with the lowest relative error. As the temperature is decreased, the error of both methods increases but RPSH acts much better than FSSH in preserving detailed balance. Here, we used the NR scheme to obtain these results following the observations in the previous section. However, it should be noted that Prezhdo and coworkers earlier used a similar model to show that VR scheme has a slightly improving effect on preserving detailed balance with FSSH in a temperature range of 350 K$-$2500 K.\cite{Sifain:2016} In Fig.~\ref{fig:fig7}, we demonstrate the relative error of the four different velocity re-scaling schemes in preserving detailed balance in RPSH. Among the four schemes, the VR scheme indeed leads to almost perfect preservation of detailed balance in both high and low temperatures. Nevertheless, regardless of the  employed velocity re-scaling scheme, RPSH always operates better than FSSH at low temperature limit. This can be traced back to the natural preservation of the quantum Boltzman distribution in RPMD which is manifested even in the multi-state dynamics here.
\begin{figure}[h!]
  \includegraphics[width=.48\textwidth]{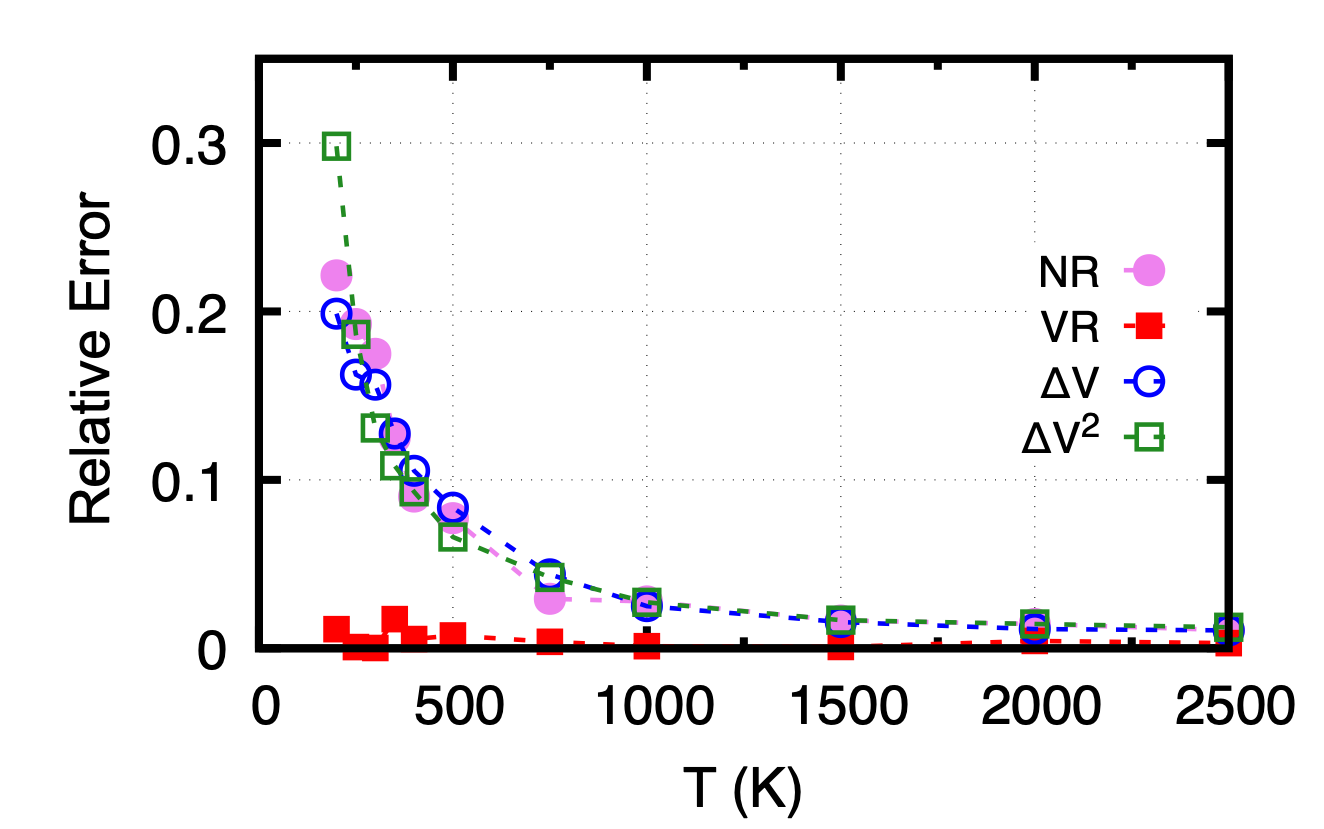}
  \caption{Relative errors as defined before in RPSH results with four different velocity re-scaling schemes.}
  \label{fig:fig7}
\end{figure}
\subsection{Differentiating between the two models}
\begin{figure}[h!]
  \includegraphics[width=.45\textwidth]{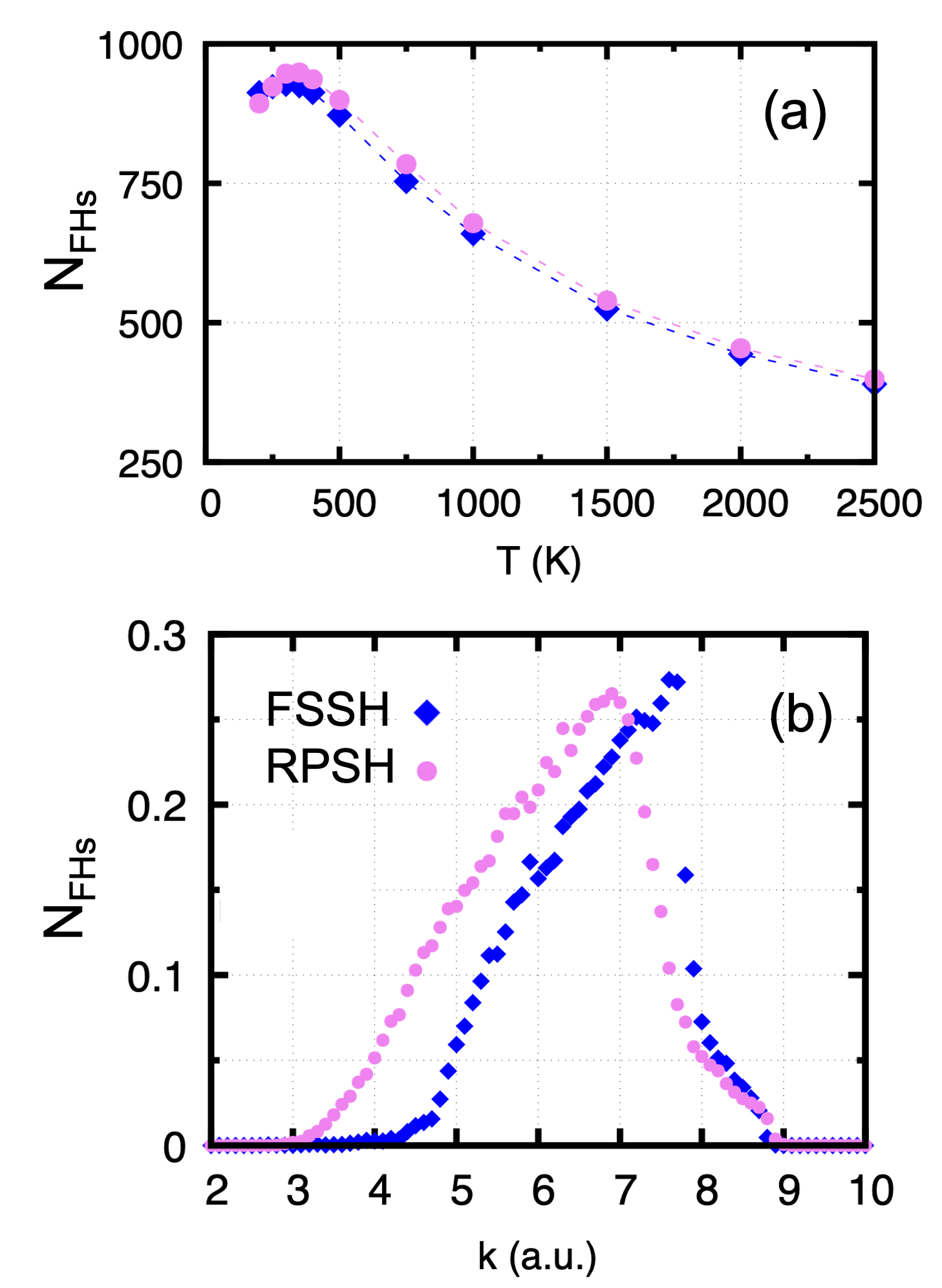}
  \caption{Number of frustrated hops averaged over number of trajectories in the chain model over a range of temperatures (a) and in the SAC model over a range of initial momentum (b) obtained from RPSH and FSSH methods.}
  \label{fig:fig6}
\end{figure}

We showed in the case of the SAC model that reversing the velocity of the adiabatic RPSH trajectories prevents an otherwise successful trajectory from going from the reactant to the product side. This observation can be generalized to any situation where tunneling through the energy barrier is a prominent factor in the success of adiabatic reactions. It should be noted that one can enforce the correct dynamics even with total or partial velocity reversing in such systems by not allowing the trajectories to attempt a transition. However, that needs \textit{a priori} information about the reaction regimes of different systems. On the other hand, regardless of the velocity re-scaling scheme, RPSH is more successful than FSSH in preserving detailed balance in the case of the 2-state quantum subsystem coupled to a linear chain of 20 nuclear DOFs. The reason can be traced back to two factors. First, RPMD technically is a phase-space representation for
the quantum Boltzmann distribution. Second, this extended phase-space, with a wider distribution of momentum, leads generally to a higher number of frustrated hops in RPSH compared to FSSH, see Fig.~\ref{fig:fig6}a. And, as we know the existence of frustrated hops is the reason behind approximate preservation of the detailed balance in surface hopping methods. It is noted that $N_{FHs}$ of the RPSH method is similarly higher than  FSSH in the case of the SAC model, Fig.~\ref{fig:fig6}b. 
Then, why the real-time dynamics in the SAC model is more accurately recovered by not reversing the velocity of the frustrated hops in RPSH trajectories while reversing it leads to exactly preserving the detailed balance in the chain model, as was already shown for the FSSH case as well
?\cite{Sifain:2016} One should note the essential differences between the two models with the lack of an avoided crossing in the chain model being the most important one. The chain model is designed to be a representative of a quantum system in thermal equilibrium with its surrounding. Furthermore, it is not suitable for investigating scattering events. Hence, the important observation here is the superiority of RPSH in preserving detailed balance with the additional insight that one can fine tune the results by choosing an appropriate treatment for frustrated hops. On the other hand, real-time dynamics of multi-state systems is best recovered by not interfering with the role of NQEs throughout the dynamics. This is of course not the case for other surface hopping methods that neglect NQEs. Nevertheless, care should be taken in the choice of an appropriate treatment of frustrated hops with respect to the system of interest.

\section{Conclusions and Outlooks}
In this work, Ring Polymer Surface Hopping (RPSH) non-adiabatic molecular dynamics methodology is critically revisited in order to shed light on its accuracy and validity limits. The most exciting aspect of RPSH, i.e., the inclusion of nuclear quantum effects into molecular dynamics, can be further harvested if the cons and pros of the method are exposed and documented in different systems and reaction regimes. 
RPSH, similar to other surface hopping algorithms, conserves the total quantum plus classical energy at the level of each individual trajectory. This leads to the creation of frustrated hops that, although they disrupt the internal consistency of the method but lead to the conservation of exact quantum Boltzmann distribution. We showed that RPSH yields better results in preserving detailed balance than the original FSSH method in the low-temperature region using a 2-state system coupled to Langevin dynamics via a linear chain of 20 nuclear DOFs. We also showed that different schemes of re-scaling velocity have a profound effect on preserving correct quantum Boltzmann populations where reversing the velocity after each frustrated hop leads to almost exact conservation of detailed balance in a range of low to high temperatures. This was opposite to what we observed in calculating the branching probabilities in Tully's avoided crossing model. While we have more frustrated hops in RPSH than FSSH for this model, we also have an interplay between frustrated hops and nuclear tunneling in the adiabatic reaction regime. Reversing the velocity of frustrated hops can disrupts the motion of an adiabatic trajectory that otherwise would tunnel through a reaction barrier from the reactant to the product side. The correct real-time dynamics in the case of this avoided crossing model is preserved better by not reversing the velocity of frustrated hops. Hence, we advise caution in choosing the proper treatment of frustrated hops in RPSH where one needs to consider the nature of the problem to be studied. 

To further bring RPSH to the mainstream, research should be directed on (i) a systematic implementation of decoherence correction and (ii) ,if necessary, reformulating the algorithm for the proper treatment of internal consistency. Surface-hopping methods take care of electronic coherence in an elegant way. However, they have always been plagued with the over-coherence problem or neglect of decoherence.\cite{Subotnik:2011} We have already shown that, in contrast to FSSH, RPSH can capture distribution-dependent decoherence in Tully’s "extended coupling with reflection" model even with a deterministic initial momentum.\cite{Shakib:2017} This improvement is because, despite initial deterministic momentum, each bead’s momentum gradually differs through the dynamical propagation, resulting in a broader centroid momentum distribution compared to the distribution in FSSH. Therefore, different reflected ring polymer trajectories gain different phases, and over an ensemble of the ring polymer trajectories, the high-frequency oscillations in reflection coefficients cancel out. However, we cannot yet claim this is a general effect. In fact, the original FSSH can also enjoy similar improvements if the initial conditions of the MD simulations involve a Maxwell-Boltzmann distribution instead of deterministic momenta. Application of RPSH in more sophisticated models designed for investigation of decoherence as well as equipping it with different decoherence correction schemes is currently being pursued in our group. On the other hand, as briefly mentioned in the methodological considerations, surface hopping methods violate internal consistency and as a result, populations of electronic states can be different if they are calculated based on the number of nuclear trajectories on each surface or the electronic amplitudes of those states or a mixture of both.\cite{Landry:2013} Recently, there has been a growing argument on the importance of dealing with this issue rather than "decoherence correction`` to systematically improve the surface hopping results.\cite{Mannouch:2023} A part of the research in our group focuses on this issue in the context of the real-time population transfer dynamics of a couple of three-state Morse potential model systems that are designed for investigating photo-excited relaxation. As can be seen in the SI Fig. S2, RPSH yields time-dependent diabatic state populations that are in good agreement with the exact quantum dynamics simulations for all three models. However, the three approaches explained earlier give different population profiles. On the other hand, according to our observations, no frustrated hops are recorded for these simulations. Hence, these models provide a unique case for disentangling the concepts of "internal consistency" and "decoherence correction" in the context of the RPSH methodology.
\begin{acknowledgments}
The authors acknowledge support from the New Jersey Institute of Technology (NJIT). This research has been enabled by the use of computing resources and technical support provided by the HPC center at NJIT. This work partially used Bridges2 at Pittsburgh Supercomputing Center through allocation CHE200007 from the Extreme Science and Engineering Discovery Environment (XSEDE), which was supported by National Science Foundation Grant no. 1548562.\cite{xsede} 
\end{acknowledgments}
\section*{Conflict of Interest}
The authors have no conflicts to disclose.
\section*{Data Availability Statement}
The data that support the findings of this study are available within the article and its supplementary material. 

\nocite{*}
\bibliography{bib}

\end{document}